\documentclass[10pt, letterpaper]{IEEEtran}

\usepackage{amsmath,amssymb,amsfonts}
\usepackage{graphicx}
\usepackage{textcomp}
\usepackage{xcolor}
\usepackage{cite}
\usepackage{algorithm}
\usepackage{algpseudocode}
\usepackage{booktabs}
\usepackage{siunitx}
\usepackage{tikz}
\usepackage{float}
\usepackage{pgfplots}
\pgfplotsset{compat=1.18}
\usepackage[hidelinks]{hyperref}   
\usepackage{multirow}
\usepackage{makecell}
\usepackage{cite}
\usepackage{hyperref,enumitem}
\usepackage{orcidlink}

\newcommand{\snr}{\mathrm{SNR}}
\newcommand{\thetamin}{\theta_{\min}}
\newcommand{\HO}{\mathrm{HO}}

\newcommand{\NHO}{N_{\HO}}

\begin{document}

\title{Multi-Criterion Target Selection for Handover in LEO In-Flight Connectivity Networks}

\author{Victor Monzon Baeza$^{_{\orcidlink{0000-0002-0035-3944}}}$,~\IEEEmembership{Senior Member,~IEEE}, and Rafael Arellano~García
\thanks{Manuscript received XXX, XX, 2026; revised XXX, XX, 2026.}
\thanks{V. M. Baeza (\textit{Corresponding author}) is with Escuela Superior de Ingeniería y Tecnología (ESIT) at Universidad Internacional de La Rioja (UNIR), Logroño, Spain. 26006. and R. A. García is with the Universitat Oberta de Catalunya (UOC), Barcelona, Spain, 08018. (e-mail: victor.monzon@unir.net; rarellano@uoc.edu).}
}

\markboth{IEEE Wireless Communications Letters,~Vol.~XX, No.~X, Sep~2026}%
{Monzon \MakeLowercase{\textit{et al.}}: Anticipatory Handover for Aeronautical In-Flight Connectivity over LEO Constellations}

\maketitle

\begin{abstract}
In-flight connectivity (IFC) over Low Earth Orbit (LEO) constellations requires frequent inter-satellite handovers whose target selection shapes the delivered quality of service. Conventional reactive schemes switch on instantaneous signal-to-noise ratio (SNR) alone, ignoring the residual visibility of the candidate satellites. This letter proposes a lightweight, terminal-local handover scheme for aeronautical LEO IFC based on a multi-criterion cost function that jointly weighs SNR, elevation, latency, spectral efficiency and an anticipatory term decaying with the residual visibility time of each candidate, regularized by a hysteresis margin that suppresses ping-pong. Evaluated by Monte Carlo simulation over $N=500$ constellation realizations against 3GPP-NTN-aligned baselines, including the event-A3 trigger with time-to-trigger, and at an identical coverage outage and the same number of handovers, the scheme improves mean spectral efficiency by up to 4.2\% and reduces latency by up to 3.1\%, both significant at 95\% confidence. It matches the 3GPP A3+TTT trigger in quality of service while keeping the handover count at the baseline level under channel noise. The benefit is therefore attributable to which satellite is selected rather than to how frequently transitions occur.
\end{abstract}


\section{Introduction}
\label{sec:intro}
Low Earth Orbit (LEO) satellite constellations have
emerged as a key enabler of broadband in-flight connectivity (IFC),
offering one-way propagation latencies of a few milliseconds and
throughput compatible with high-quality passenger service over
oceanic and remote routes \cite{parada2026integrated}.
Their architectural appeal, however, comes at a structural cost:
the high orbital speed of LEO satellites confines individual
visibility windows of an aircraft to a few minutes, forcing the
on-board terminal to perform frequent inter-satellite handovers
(HOs) to sustain the link. The aeronautical regime is especially
demanding in this respect, since aircraft velocity, narrow useful
elevation ranges, and the absence of stable terrestrial fallback
amplify the operational impact of every mistimed
transition~\cite{park2021trends}.
 
Standard mobility procedures in 3GPP non-terrestrial networks (NTN)
remain measurement-based and reactive in nature: the HO is triggered
when the link quality of the serving satellite falls below a
configured threshold, and the target is selected from the visible
set as the one offering the best instantaneous signal-to-noise ratio
(SNR)~\cite{3gpp38821,juan2022performance}. Such reactive policies
inherit two well-known limitations in the LEO regime: they let the
link degrade before acting, accumulating outage and packet loss;
and they ignore the residual visibility horizon of the target,
occasionally selecting satellites that are themselves about to
disappear.
 
Several recent works exploit the deterministic orbital motion to plan HOs in advance: graph-based shortest-path search~\cite{hozayen2022graph}, the offline ephemeris-dependent planner PreHO~\cite{he2026preho}, and the network-side, ML-based PHandover~\cite{wu2025phandover}, the latter evaluated for stationary ground UEs. Multi-criterion triggers that fold remaining visible or service time into the decision have also been studied, jointly with link quality and satellite load~\cite{shi2022load} or within a learned utility of several QoS factors~\cite{li2026mappo}, yet these rely on network-side information, load state, or trained policies. In contrast, ours is online and terminal-local: the decision uses the locally observable $\{\theta_i(t),d_i(t),\tau_i(t)\}$ alone, no global ephemeris, network function, or trained component, with a closed-form anticipatory term; unlike prior visibility-aware triggers, we further isolate its effect under a common, scheme-agnostic outage definition, so the reported gains reflect target selection alone. Table~I positions it against these works; to our knowledge, no such scheme for a moving aircraft exists in the open literature.
\begin{table}[!ht]
\centering
\caption{Qualitative positioning of the proposed scheme against the reactive baseline and the closest anticipatory and visibility-aware handover works.}
\label{tab:sota}
\setlength{\tabcolsep}{4pt}
\renewcommand{\arraystretch}{1.15}
\footnotesize
\begin{tabular}{@{}lllll@{}}
\toprule
Method & Decision locus & Mode & \makecell[l]{Anticipation\\mechanism} & \makecell[l]{UE\\mobility} \\
\midrule
\makecell[l]{Reactive\\\,[5],[6]} & \makecell[l]{On-board\\terminal} & Online & \makecell[l]{None\\(reactive trigger)} & Generic \\
\makecell[l]{Hozayen\\\emph{et al.} [7]} & \makecell[l]{Central\\planner} & Offline & \makecell[l]{Graph\\shortest-path} & \makecell[l]{Generic\\LEO} \\
PreHO [8] & \makecell[l]{Central\\planner} & Offline & \makecell[l]{Alternating\\optimization} & \makecell[l]{Generic\\LEO} \\
PHandover [9] & \makecell[l]{Network side\\(5G core)} & Online & \makecell[l]{ML signal\\predictor} & \makecell[l]{Stationary\\ground UE} \\
\makecell[l]{Shi\\\emph{et al.} \cite{shi2022load}} & \makecell[l]{Network side\\(load-aware)} & Online & \makecell[l]{Remaining\\visible time} & \makecell[l]{Generic\\LEO} \\
\makecell[l]{Li\\\emph{et al.} \cite{li2026mappo}} & \makecell[l]{Network side\\(MARL)} & Online & \makecell[l]{Learned QoS\\utility} & \makecell[l]{Ground\\(STIN)} \\
\textbf{Proposed} & \makecell[l]{\textbf{On-board}\\\textbf{terminal}} & \textbf{Online} & \makecell[l]{\textbf{Closed-form}\\$\boldsymbol{e^{-\alpha\tau_i}}$} & \makecell[l]{\textbf{Moving}\\\textbf{aircraft}} \\
\bottomrule
\end{tabular}
\end{table}

This letter contributes: (i) a five-component multi-criterion cost function with a closed-form anticipatory term $e^{-\alpha\tau_i(t)}$ and a hysteresis margin $\Delta$ that decouples responsiveness from ping-pong under stochastic SNR; (ii) a Monte Carlo evaluation with a common, scheme-agnostic outage definition, isolating the effect of \emph{target selection} from coverage geometry; and (iii) an ablation and comparison against stronger baselines (A3+TTT, $\tau$-aware) showing that the spectral-efficiency and latency gains arise from proactive target selection, while the multi-criterion and hysteresis keep the handover count stable under channel noise, matching the 3GPP A3+TTT trigger at a lower handover rate.


\begin{figure}[!ht]
    \centering
    \includegraphics[width=1\linewidth]{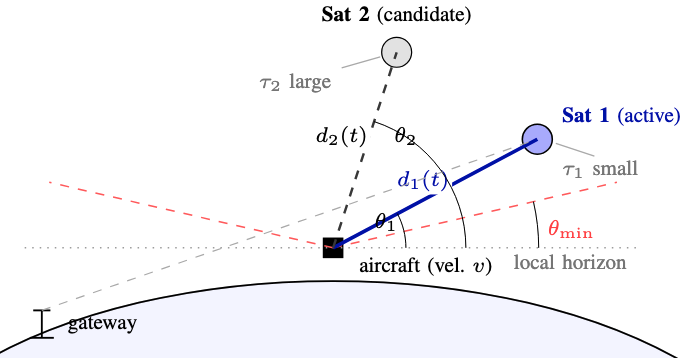}
    \caption{Aeronautical IFC over an LEO constellation. The on-board terminal observes for each visible satellite $i$ the elevation $\theta_i(t)$, the slant distance $d_i(t)$, and the residual visibility time $\tau_i(t)$. The active satellite (solid link) is approaching the minimum-elevation horizon $\thetamin$ with small $\tau_1(t)$; the candidate (dashed link) has just entered the visible set with large $\tau_2(t)$. The handover decision is taken locally by minimising the cost $C_i(t)$ defined in~\eqref{eq:cost}.}
    \label{fig:system}
\end{figure}

\section{System Model}
\label{sec:model}
We consider a single-aircraft IFC terminal accessing a LEO constellation of $M$ satellites along a flight segment of duration $T$, sampled at discrete time instants $t \in \{0, \Delta t, \dots, T\}$, as illustrated in Fig.~\ref{fig:system}. The decision logic is executed locally on the on-board terminal, using only information available at that node.
 
\subsection{Geometry and Visibility}
\label{ssec:geometry}
At each instant $t$, satellite $i$ is characterized by its elevation angle $\theta_i(t)$ above the local horizon and its slant distance $d_i(t)$ to the aircraft. A satellite is declared visible when $\theta_i(t) \ge \thetamin$, with $\thetamin$ a configurable operational threshold; the visible set is denoted by $\mathcal{S}(t) \subseteq \{1, \ldots, M\}$. For each visible satellite, the residual visibility time $\tau_i(t)$ is the time remaining until $\theta_i$ falls below $\thetamin$, computed from the satellite trajectory and the aircraft position. In practice, $\tau_i(t)$ need not be signaled by the network: since the aircraft trajectory is smooth over a handover horizon and the orbital motion is deterministic, the on-board terminal estimates it locally, either by propagating the satellite ephemeris carried in broadcast two-line elements (TLE) and solving $\theta_i(t+\tau_i)=\theta_{\min}$, or, in the ephemeris-free case, by short-horizon extrapolation of the locally observed elevation track $\theta_i(t)$. Either route keeps the decision terminal-local and consistent with the autonomy and low-latency requirements of aeronautical operation. The constellation comprises $M=8$ satellites, each described by a trapezoidal elevation pass with parameters sampled independently: maximum elevation $\theta_{i}^{\max}\!\in\![30^\circ, 80^\circ]$, pass duration $T_{i}^{\mathrm{pass}}\!\in\![8, 15]$\, min, and a pass offset that yields one or two satellites simultaneously in view over the simulated hour. This abstraction is consistent with the operating regime of dense LEO constellations such as Starlink at mid-latitudes, where the on-board terminal typically sees one or two satellites simultaneously above $10^\circ$ elevation, and is deliberately decoupled from a specific ephemeris to isolate the behavior of the decision logic from constellation-design choices. The algorithm presented below operates on $\{\theta_i(t), d_i(t), \tau_i(t)\}$ as inputs, and therefore applies unchanged when these quantities are obtained from real TLE data of a specific constellation.
 
\subsection{Link Model}
\label{ssec:link}
The aircraft--satellite link is dominated by line-of-sight (LoS)
propagation and is well described by a free-space path-loss
model~\cite{baeza2022channel}. Denoting the transmit power by
$P_t$, the antenna gains by $G_t, G_r$, the carrier wavelength by
$\lambda$ and the equivalent noise power by $N_0 B$ over a
bandwidth $B$, the received SNR on the link to satellite $i$ is
\begin{equation}
\label{eq:snr}
\snr_i(t)\;=\;\frac{P_t\,G_t\,G_r\,\lambda^2}
                  {(4\pi)^2\,d_i^2(t)\,N_0 B},
\end{equation}
which yields the proportionality $\snr_i(t) \propto 1/d_i^2(t)$
under the assumption of constant terminal and noise parameters.
For the purposes of the comparative study, the link-budget
constants are absorbed into a per-satellite envelope and only the
$\snr_i(t)$ profile relative to the maximum elevation of each pass
is retained; the variability scenario adds a zero-mean Gaussian
perturbation on top of this envelope.
The one-way propagation latency to satellite $i$ is
\begin{equation}
\label{eq:lat}
\ell_i(t)\;=\;d_i(t)/c,
\end{equation}
with $c$ the speed of light, and the achievable spectral efficiency is taken as
\begin{equation}
\eta_i(t) = \log_2\!\big(1 + \mathrm{SNR}_i(t)\big).
\label{eq:eta}
\end{equation}
Each executed handover incurs a fixed penalty of $T_\HO$ seconds during which the link is unavailable; this lumped model captures the aggregate cost of the standard random access channel (RACH)-based handover procedure without resolving its individual signaling steps. A handover failure is recorded whenever no visible satellite is available at the time of execution. This visibility-based definition isolates the geometric failure mode addressed by the anticipatory term; other practical failure modes, such as RACH or signaling failures and synchronization loss during the transition, are abstracted into the lumped penalty $T_{\mathrm{HO}}$ and are not resolved individually here.
 
\subsection{Multi-Criterion Cost Function}
\label{ssec:cost}
The link quality offered by each visible candidate is summarized by
a scalar cost
\begin{equation}
\label{eq:cost}
C_i(t) \;=\; \sum_{k=1}^{5} w_k\,\tilde{f}_{k,i}(t),
\qquad
\sum_{k=1}^{5} w_k = 1,
\quad w_k \ge 0,
\end{equation}
where the five normalized components $\tilde{f}_{k,i}(t) \in [0,1]$
capture, respectively:
(i) link quality from SNR;
(ii) geometric margin from elevation angle;
(iii) propagation latency from slant distance;
(iv) achievable spectral efficiency $\eta_i(t)$; and
(v) the \emph{anticipatory failure risk}
\begin{equation}
\label{eq:ffail}
\tilde{f}_{\mathrm{fail},i}(t) \;=\; e^{-\alpha\,\tau_i(t)},
\qquad \alpha > 0.
\end{equation}
Components (i), (ii) and (iv) are inverted before normalization so
that larger values correspond to higher cost; min--max
normalization is applied across the visible set at each $t$. All
the SNR-, latency- and efficiency-based terms are instantaneous;
the novelty resides in~\eqref{eq:ffail}, which converts the residual
visibility time into a smooth penalty that becomes dominant only as
$\tau_i(t) \to 0$. 
he scheme's forward-looking component is ~\eqref{eq:ffail}. Unlike the instantaneous components (i)-(iv) in~\eqref{eq:cost}, which score only the present link state, $\tilde{f}_{\mathrm{fail},i}(t)$ scores the \emph{near-future} availability of the link: it grows as a candidate approaches the end of its visibility window, so a satellite that is excellent at the current instant but about to set is penalised \emph{before} its quality actually collapses. This is the sole mechanism that renders the decision anticipatory rather than reactive. The exponential form is adopted because it keeps the penalty negligible while $\tau_i(t)$ is large and makes it rise smoothly and monotonically as $\tau_i(t)\!\to\!0$, in contrast with a hard residual-time threshold, which would inject a discontinuity into the cost and reintroduce the late, abrupt switching the scheme is meant to avoid. Its sensitivity admits a closed form, 
\begin{equation}
\frac{\partial \tilde{f}_{\mathrm{fail},i}}{\partial \tau_i} = -\,\alpha\, e^{-\alpha\tau_i(t)},
\label{eq:sensitivity}
\end{equation}
whose magnitude is maximal exactly as $\tau_i(t)\!\to\!0$, i.e.\ the penalty reacts most strongly precisely where a mistimed transition is most costly. The decay constant $\alpha$ thus fixes a characteristic anticipation horizon $1/\alpha$: candidates with $\tau_i(t)\!\gg\!1/\alpha$ are treated as effectively persistent, whereas those with $\tau_i(t)\!\lesssim\!1/\alpha$ are progressively down-ranked. In the limits, $\alpha\!\to\!0$ recovers a purely instantaneous score (reactive limit) and $\alpha\!\to\!\infty$ ranks candidates by residual visibility time alone. The reference value $\alpha=0.05$ places the anticipation horizon in the intermediate region that trades foresight against premature switching, as the sensitivity analysis of Sec.~\ref{sec:results} confirms.
In what follows, we adopt equal weights $w_k = 0.2$ as a neutral baseline that lets the selection policy, rather than weight tuning, drive the comparison.

\section{Handover Decision Rule}
\label{sec:algo}
\subsection{Multi-Criterion Target Selection}
\label{ssec:predictive}
At every instant $t$, the terminal evaluates the cost $C_i(t)$ defined in~\eqref{eq:cost} for every visible satellite $i \in \mathcal{S}(t)$ and identifies the best candidate as
\begin{equation}
\label{eq:argmin}
s^\star(t) \;=\; \arg\!\min_{i \in \mathcal{S}(t)} C_i(t).
\end{equation}
A handover from the active satellite $s(t-1)$ to $s^\star(t)$ is executed only when the cost improvement exceeds a hysteresis margin $\Delta \ge 0$:
\begin{equation}
\label{eq:trigger}
s(t) \;=\;
\begin{cases}
s^\star(t) & \text{if } C_{s^\star(t)}(t) + \Delta < C_{s(t-1)}(t), \\
s(t-1)     & \text{otherwise.}
\end{cases}
\end{equation}
When~\eqref{eq:trigger} fires, the link is unavailable during a penalty interval of duration $T_\HO$. If $\mathcal{S}(t) = \emptyset$, no transition is possible and the terminal registers a service outage. The decision rule in~\eqref{eq:argmin}-\eqref{eq:trigger} thus has constant complexity per visible satellite and operates exclusively on information available at the on-board terminal, requiring no network-side coordination. 

\subsection{Reactive Baseline}
\label{ssec:reactive}
The baseline against which~\eqref{eq:trigger} is compared follows the measurement-based, threshold-driven structure adopted by 3GPP NTN mobility procedures~\cite{3gpp38821,juan2022performance}. A handover is triggered whenever
\begin{equation}
\label{eq:reactive}
\snr_{s(t-1)}(t) < \gamma_{\min}
\quad \text{or} \quad
\theta_{s(t-1)}(t) \le \thetamin,
\end{equation}
and the target is selected as $s(t) = \arg\max_{i\in \mathcal{S}(t)\setminus\{s(t-1)\}} \snr_i(t)$; if $\mathcal{S}(t) = \emptyset$ the terminal records a service outage. The threshold $\gamma_{\min}$ plays here the role of the cell-quality threshold employed by the 3GPP NR measurement-based handover procedure; we set $\gamma_{\min}$ to the value that maximizes the spectral efficiency of the baseline, ensuring a fair comparison. The same penalty $T_\HO$ is applied per transition. The baseline thus ignores both the residual visibility time of the candidates and any criterion beyond instantaneous SNR; it also lacks an explicit hysteresis margin, mirroring the simplest NTN handover trigger before the conditional-handover enhancements of 3GPP Release 17. Any performance difference in Section~\ref{sec:results} stems from the proactive multi-criterion selection of~\eqref{eq:cost}-\eqref{eq:trigger} rather than from the reactive trigger of~\eqref{eq:reactive}.

\section{Numerical Results}
\label{sec:results}
\begin{figure}[!ht]
\centering
\includegraphics[width=\columnwidth]{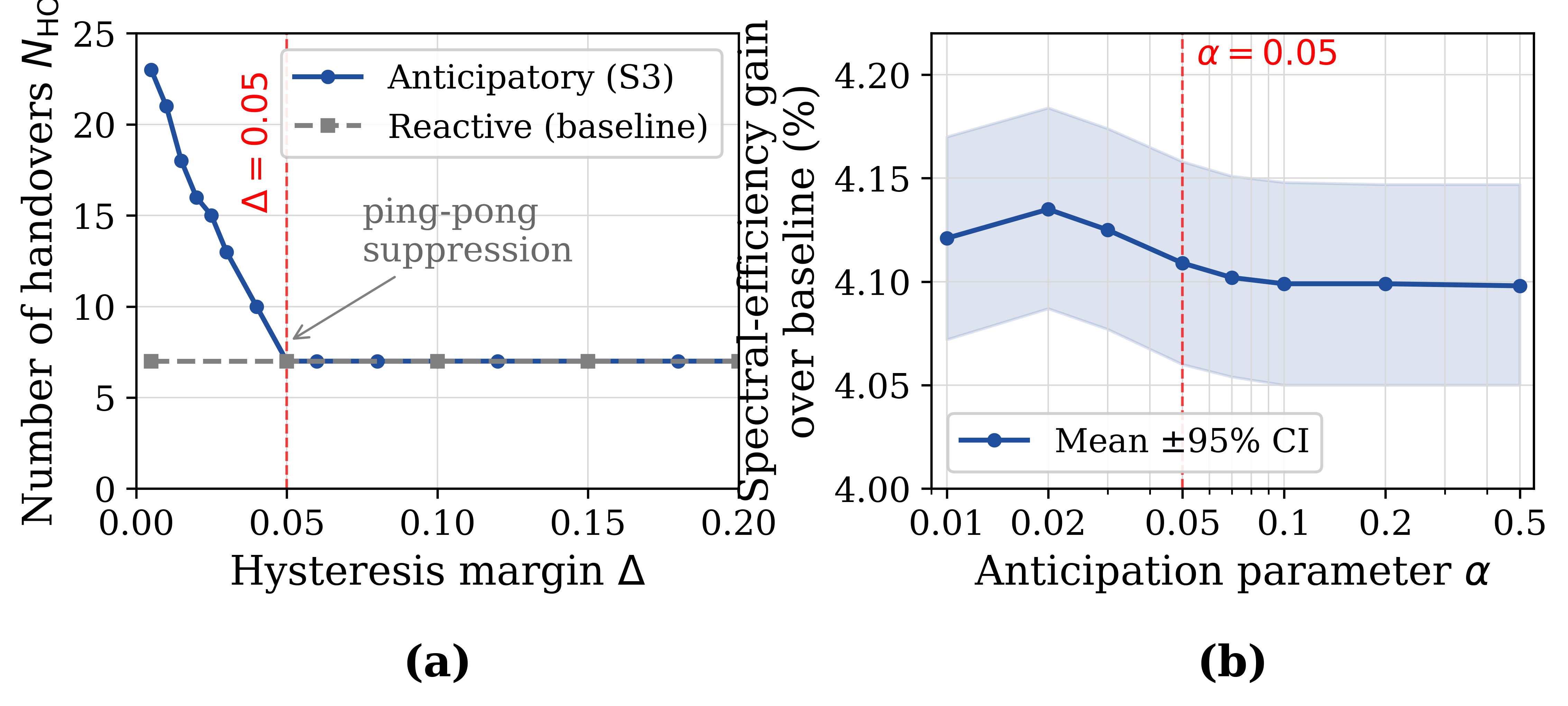}
\caption{Sensitivity over $N=500$ realizations: (a)~$\NHO$ vs.\ $\Delta$ (S$_3$); (b)~spectral-efficiency gain vs.\ $\alpha$ (mean $\pm$95\% CI).}
\label{fig:sensitivity}
\end{figure}

The proposed scheme is compared against the reactive baseline and the stronger baselines over a $T = 60$~min flight segment with sampling step $\Delta t = 1$~s. The proposed scheme uses $\alpha=\Delta=0.05$ and equal weights $w_k=0.2$; these reference values are not fitted to the metrics: $\alpha$ provides a moderate anticipation horizon $1/\alpha$ that balances sensitivity to imminent visibility loss against premature switching, while $\Delta$ is set to the smallest margin that suppresses noise-induced ping-pong, as established in Fig.~\ref{fig:sensitivity}(a); the reactive threshold $\gamma_{\min}$ is set to the value that maximizes the baseline spectral efficiency, ensuring a fair comparison. Each configuration is run on the same synthetic constellation realization, so the only source of variation between schemes is the decision logic itself. 
The simulation parameters are: $M=8$ satellites, $\theta_{\min}=10^\circ$, $\gamma_{\min}=0.15$, per-transition penalty $T_{\mathrm{HO}}=1$~s (3~s in S$_2$), and $N=500$ Monte Carlo realizations. Aircraft motion is not parameterized by an explicit speed: the combined satellite--aircraft geometry is captured by the per-pass elevation profiles (maximum elevation $\theta_i^{\max}$ and pass duration $T_i^{\mathrm{pass}}$), so the results depend on the relative dynamics rather than on a specific cruise velocity.
Four operating regimes are defined in Table~\ref{tab:scenarios}: three representative scenarios in which the proposed scheme is evaluated against the baseline (S$_1$--S$_3$), and a fourth high-elevation regime (S$_4$) that delimits the applicability boundary of the anticipatory term, as discussed below.

\begin{table}[!ht]
\centering
\caption{Definition of the simulation scenarios. All scenarios share
$M = 8$ satellites, $T = 60$~min, and $\Delta t = 1$~s.}
\label{tab:scenarios}
\renewcommand{\arraystretch}{1.15}
\setlength{\tabcolsep}{5pt}
\footnotesize
\begin{tabular}{@{}clccl@{}}
\toprule
ID & Description     & $\thetamin$ & $T_\HO$~(s)
   & SNR perturbation \\
\midrule
S$_1$ & Nominal               & $10^\circ$ & 1 & none \\
S$_2$ & Higher HO penalty     & $10^\circ$ & 3 & none \\
S$_3$ & Stochastic SNR        & $10^\circ$ & 1
      & Gaussian, $\sigma_{\snr}\!=\!0.05$ \\
S$_4$ & Adverse geometry      & $20^\circ$ & 1 & none \\
\bottomrule
\end{tabular}
\end{table}

We report the coverage outage $T_{\mathrm{cov}}$, accumulated over instants with no visible satellite (purely geometric, hence identical for both schemes), together with mean spectral efficiency $\bar{\eta}$ and latency; a service-interruption time measured with a common usability threshold is used only to verify that availability is equal across schemes.

Table~\ref{tab:results} reports means over $N=500$ realizations with 95\% CIs and confirms three observations. \textit{First}, availability is equal for both schemes: the coverage outage $T_{\mathrm{cov}}$ is purely geometric and identical ($39.1\pm5.0$~s; paired difference $0$), and the service-interruption time, measured with a common usability threshold, is statistically indistinguishable between schemes. Both policies therefore deliver the same outage, which isolates the effect of \emph{target selection} from coverage geometry. \textit{Second}, at this equal outage and an unchanged number of handovers ($\NHO\approx 7$; paired difference $+0.01$, below $0.2\%$), the proposed scheme improves mean spectral efficiency by $+2.9\%$ to $+4.2\%$ and reduces latency by $-1.9\%$ to $-3.1\%$ across scenarios, all significant at 95\% confidence. \textit{Third}, Table~\ref{tab:ablation_baselines} isolates the source of the gain and compares against stronger baselines. Removing the anticipatory term, the multi-criterion (SNR-only cost), or the hysteresis leaves the spectral-efficiency gain essentially unchanged, so the gain stems from \emph{proactive} best-candidate selection rather than from any single component: continuously minimizing the cost keeps the terminal on a better satellite than the threshold-triggered baseline, which commits until the serving link degrades. The ingredients instead govern stability: under stochastic SNR, removing the hysteresis or the multi-criterion inflates $\NHO$ from $7$ to $20$ and $16$, respectively, while the anticipatory term is a minor stabilizer. Against stronger baselines, a visibility-aware reactive heuristic ($\tau$-aware) barely improves on the reactive baseline ($+0.14\%$), confirming that anticipating visibility loss is not by itself the source of the gain, whereas the canonical 3GPP event-A3 trigger with time-to-trigger (A3+TTT) matches the proposed scheme in spectral efficiency; however, its SNR-only trigger reacts to noise and inflates $\NHO$ to $7.6$ against $7.0$. The proposed scheme therefore attains the same QoS as the standard 3GPP trigger at a stable, baseline-level handover count, with a closed-form, terminal-local rule and no per-event offset or timer tuning. We note that a threshold baseline additionally declares failure when a degraded serving link has no visible alternative, an instant the proposed scheme keeps serving at reduced quality; we do not attribute this policy difference to the anticipatory mechanism and report the geometry-bound $T_{\mathrm{cov}}$ as the common availability measure.
In the adverse-geometry regime S$_4$ ($\thetamin=20^\circ$), coverage is structurally limited and the schemes coincide in outage, while the spectral-efficiency and latency gains persist.

\begin{table}[!ht]
\centering
\caption{Performance over $N=500$ constellation realizations ($\pm$95\% CI). Coverage outage $T_{\mathrm{cov}}$ is purely geometric and identical for both schemes; the gains appear in spectral efficiency and latency at an unchanged number of handovers.}
\label{tab:results}
\setlength{\tabcolsep}{4pt}
\begin{tabular}{llccc}
\hline
Scenario & Scheme & $T_{\mathrm{cov}}$ (s) & $\bar{\eta}$ (b/s/Hz) & $\bar{d}/c$ (ms) \\
\hline
\multirow{2}{*}{\makecell[l]{S1\\Nominal}}
 & Reactive   & $39.1\!\pm\!5.0$ & $0.683\!\pm\!0.005$ & $2.434\!\pm\!0.012$ \\
 & Anticipatory & $39.1\!\pm\!5.0$ & $0.711\!\pm\!0.005$ & $2.358\!\pm\!0.012$ \\
\hline
\multirow{2}{*}{\makecell[l]{S2\\($T_{\mathrm{HO}}{=}3$\,s)}}
 & Reactive   & $39.1\!\pm\!5.0$ & $0.680\!\pm\!0.005$ & $2.434\!\pm\!0.012$ \\
 & Anticipatory & $39.1\!\pm\!5.0$ & $0.709\!\pm\!0.005$ & $2.358\!\pm\!0.012$ \\
\hline
\multirow{2}{*}{\makecell[l]{S3\\($\sigma_{\mathrm{SNR}}{=}0.05$)}}
 & Reactive   & $39.1\!\pm\!5.0$ & $0.690\!\pm\!0.005$ & $2.403\!\pm\!0.012$ \\
 & Anticipatory & $39.1\!\pm\!5.0$ & $0.710\!\pm\!0.005$ & $2.358\!\pm\!0.012$ \\
\hline
\end{tabular}
\end{table}

\begin{table}[!ht]
\centering
\caption{Ablation and baseline comparison over $N=500$ realizations (S3, stochastic SNR): spectral-efficiency gain over the reactive baseline and handover count. All methods share the same coverage outage ($T_{\mathrm{cov}}=39.1$~s). Removing an ingredient barely changes the gain but strongly affects $\NHO$; the proposed scheme matches the 3GPP A3+TTT trigger at a lower, baseline-level $\NHO$.}
\label{tab:ablation_baselines}
\setlength{\tabcolsep}{5pt}
\renewcommand{\arraystretch}{1.1}
\begin{tabular}{lcc}
\hline
Method / variant & $\Delta\bar{\eta}$ (\%) & $\NHO$ \\
\hline
Reactive [5],[6]                    & $+0.00$ & $7.0$ \\
$\tau$-aware                        & $+0.14$ & $7.0$ \\
A3\,+\,TTT                          & $+2.86$ & $7.6$ \\
\hline
Proposed (full)                     & $+2.89$ & $7.0$ \\
\quad $-$ anticipation ($w_5{=}0$)  & $+2.89$ & $7.3$ \\
\quad $-$ multi-criterion (SNR-only)& $+2.71$ & $15.9$ \\
\quad $-$ hysteresis ($\Delta{=}0$) & $+2.63$ & $19.7$ \\
\hline
\end{tabular}
\end{table}

The role of the hysteresis margin is isolated in Fig.~\ref{fig:sensitivity}(a). Under stochastic SNR variability, an insufficient $\Delta$ allows transient noise excursions to flip the $\arg\min$ in~\eqref{eq:argmin}, triggering up to 23 handovers in a single realization in the absence of hysteresis; from $\Delta \approx 0.05$ onward, $\NHO$ stabilizes at the value obtained by the reactive baseline, while the spectral-efficiency and latency gains reported in Table~\ref{tab:results} are preserved. Fig.~\ref{fig:sensitivity}(b) reports the Monte Carlo sensitivity to $\alpha$: the spectral-efficiency gain over the baseline is robust, holding at $\approx 4.1\%$ across two orders of magnitude ($0.01$--$0.5$) with only a shallow decline for large $\alpha$ as premature switching sets in, while $\NHO$ stays invariant. The reference $\alpha=0.05$ thus lies on a broad plateau, a non-critical operating point; together with panel~(a) this confirms the complementary split anticipated in Sec.~\ref{sec:model}: $\alpha$ governs anticipation timing, $\Delta$ governs stability.

First, the equal weights $w_k=0.2$ are a neutral rather than a tuned choice: across $100$ weight vectors sampled uniformly on the simplex, the spectral-efficiency gain stays within a 0.06-point band (within $\pm0.03$ points of the equal-weight value in S$_1$), and profiles spanning efficiency-, latency- and stability-oriented biases leave both the gain and the latency essentially unchanged, so the result is insensitive to the weighting. The weights instead affect handover stability: an efficiency-heavy profile raises $\NHO$ to $10.6$ under stochastic SNR by chasing the noisy SNR, whereas equal, latency- or stability-oriented weights keep $\NHO$ at the baseline level. Equal weights are therefore a robust default that a deployment may bias per service without materially changing the gain. Second, the link model of Sec.~\ref{sec:model} captures free-space LoS attenuation but omits Rician fading, ITU-R~P.618 gaseous/rain attenuation and elevation-dependent noise temperature, material for L/S/Ka NGSO links \cite{baeza2022channel}. As these add zero-mean impairments over a geometric quantity $\tau_i(t)$, the \emph{timing} of the transitions is largely channel-independent, whereas the SNR-dependent spectral efficiency and latency margins should be read as an upper envelope. Integration with a calibrated NGSO channel and 3GPP NTN conditional-handover signaling is ongoing and is the natural path to an ephemeris-driven validation.

\section{Conclusion}
\label{sec:conclusion}
This letter proposed a lightweight, terminal-local anticipatory handover scheme for aeronautical LEO IFC, combining a multi-criterion cost function with a closed-form residual-visibility term and a hysteresis margin. Under a common outage definition and against 3GPP-NTN-aligned baselines including A3+TTT, and at equal coverage outage and handover count, it improves mean spectral efficiency by up to 4.2\% and reduces latency by up to 3.1\% (95\% confidence). An ablation attributes these gains to proactive target selection, with the multi-criterion cost and hysteresis keeping the handover count at the baseline level under noise; the scheme thus matches A3+TTT in QoS with no per-event tuning. Validation on ephemeris-driven constellations, calibrated NGSO channels, and 3GPP NTN conditional handover is ongoing.
\bibliographystyle{IEEEtran}
\bibliography{references}

\end{document}